\documentclass[preprint,letterpaper,showpacs]{revtex4}
\usepackage{graphicx}% Include figure files
\def \re {{\rm Re}\,}
\def \im {{\rm Im}\,}

\def \vS {{\bf S}}

\def \vk {{\bf k}}

\def \ds {\displaystyle}

\def \al {\eta }
\def \bal {\bar{\eta }}

\def \hm {{\bf m}}

\def \Ts {{\rm Tr}}
\def \TC {T_{\rm C}}
\def \TN {T_{\rm N}}
\def \TM {T_{\rm SRO}}
\def \TR {T^{(\rm max )}_{\rm SRO}}
\def \Etmin {E^{({\rm min })}}
\def \E2min {E_2^{({\rm min })}}

\def \bs {{\bf \sigma }}
\def \vS {{\bf S}}

\def \seff {S_{\rm eff}}
\def \uI {\underline{I}}
\def \uGr {\underline{G}}

\def \vs {{\bf s}}

\def \bc {\bar c}
\def \JH {J_{\rm H}}
\def \tJH {J}

\begin{document}

%\preprint{\today}
\title{Short-Range Ordered Phase of the Double-Exchange Model in Infinite Dimensions}
\author{Randy S. Fishman$^*$, Florentin Popescu$^{\circ }$, Gonzalo Alvarez$^{\bullet}$,
Thomas Maier$^{\bullet}$, and Juana Moreno$^{\dagger }$}
\affiliation{$^*$Condensed Matter Sciences Division, Oak Ridge National Laboratory, Oak Ridge, TN 37831-6032}
\affiliation{$^{\circ }$Physics Department, Florida State University, Tallahassee,FL, 32306}
\affiliation{$^{\bullet}$Computer Science and Mathematics Division, Oak Ridge National Laboratory, Oak Ridge, TN 37831-6032}
\affiliation{$^{\dagger }$Physics Department, University of North Dakota, Grand Forks, ND 58202-7129}

\begin{abstract}
Using dynamical mean-field theory, we have evaluated the magnetic
instabilities and $T=0$ phase diagram of the double-exchange model
on a Bethe lattice in infinite dimensions.  In addition to
ferromagnetic (FM) and antiferromagnetic (AF) phases, we also study
a class of disordered phases with magnetic short-range order (SRO).
In the weak-coupling limit, a SRO phase has a higher transition
temperature than the AF phase for all fillings $p$ below 1 and can
even have a higher transition temperature than the FM phase.  At
$T=0$ and for small Hund's coupling $\JH $, a SRO state has lower
energy than either the FM or AF phases for $0.26 \le p < 1$.  Phase
separation is absent in the $\JH \rightarrow 0$ limit but appears
for any non-zero value of $\JH $.
\end{abstract}

\maketitle

Due to the keen interest in manganites \cite{mang} and dilute magnetic semiconductors \cite{dms},
the double-exchange (DE) model remains one of the central models in condensed-matter science.
Yet the behavior of the DE model for small Hund's coupling continues to puzzle
researchers.   Both the nature of the uniform phases and the presence of phase separation
(PS) in the weak-coupling regime is the subject of ongoing debate \cite{yun:98,nag:00,cha:00,aus:01}.
This paper uses dynamical mean-field theory (DMFT) to obtain the magnetic instabilities
and $T=0$ phase diagram of the DE model in infinite dimensions.   We reach the
surprising conclusion that a short-range ordered (SRO) phase is stable for small Hund's coupling $\JH $
and even has a higher transition temperature than the long-range ordered ferromagnetic (FM)
and antiferromagnetic (AF) phases for some fillings.

Developed in the late 1980's by M\"uller-Hartmann \cite{mul:89} and Metzner and Vollhardt \cite{met:89},
DMFT exploits the momentum independence of the self-energy in infinite dimensions.
Even in three dimensions, DMFT is believed to capture the physics of correlated systems
including the narrowing of electronic bands and the Mott-Hubbard transition \cite{geo:96}.
Although DMFT has been widely applied to the DE model
\cite{yun:98,nag:00,cha:00,aus:01,fur:95,mil:96,fis:03,che:03,fis:05},
until now there has been no complete treatment of the magnetic instabilities and $T=0$
phase diagram of the DE model.  Usually, DMFT calculations are performed on
either a Bethe or hypercubic lattice. Because of the simplified formalism on the Bethe lattice and the
risk of pathological results \cite{che:03} associated with the unbounded, hypercubic
density-of-states (DOS), we choose to work on a Bethe lattice with a semi-circular
DOS bounded by $\pm W/2$.

The DE Hamiltonian is
\begin{equation}
\label{ham}
H=-t\sum_{\langle i,j \rangle }\Bigl( c^{\dagger }_{i\alpha }c^{\, }_{j\alpha }
+c^{\dagger }_{j\alpha }c^{\, }_{i\alpha } \Bigr) -2 \JH \sum_i \vs_i \cdot \vS_i
\end{equation}
where $c^{\dagger }_{i\alpha }$ and $c_{i\alpha }$ are the creation and destruction operators
for an electron with spin $\alpha $ at site $i$,
$\vs_i =(1/2) c^{\dagger }_{i\alpha } \bs_{\alpha \beta } c^{\, }_{i\beta }$
is the electronic spin, and $\vS_i=S\hm_i $ is the spin of the local moment (treated classically).
Repeated spin indices are summed.   Then within DMFT, the effective action on site 0 is given by \cite{fur:95}
\begin{equation}
\label{act}
\seff (\hm) =-T\sum_n\bc_{0\alpha }(i\nu_n) \Bigl\{ G_0(i\nu_n)^{-1}_{\alpha \beta }
+ \tJH \bs_{\alpha \beta }\cdot \hm\Bigr\} c_{0\beta }(i\nu_n),
\end{equation}
where $\tJH =\JH S$, $\nu_n=(2n+1)\pi T$, $\bc_{0\alpha }(i\nu_n)$ and $c_{0\alpha }(i\nu_n)$ are
now anticommuting Grassman variables, and $\uGr_0(i\nu_n)$ is the bare Green's function containing dynamical
information about the hopping of electrons from other sites onto site 0.

In the infinite-dimensional limit, the high-temperature non-magnetic (NM) phases of the
Heisenberg and DE models have a vanishing correlation length $\xi $.
The SRO phase is a bulk solution of the DE model that, like the FM and AF states,
relies only on the local topology of the Bethe lattice.  It has some of the same characteristics as
conventional spin glasses:  a finite local magnetization and spin-spin correlations
that decay exponentially over distance \cite{bin:86}.  The SRO phase is characterized by a correlation parameter $q$
that gives the average $q=\langle \sin^2 (\theta_i/2) \rangle $, where $\theta_i $ is the angle between a central
spin and each neighboring spin.  Overall, the neighboring spins describe a cone with angle
$2\arcsin (\sqrt{q})$ around the central spin.  The FM and AF phases have, respectively, $q=0$ and 1.
If $M_n$ is the average magnetization of the lattice sites at a distance $na$ ($a$ is the lattice constant)
from the central site, then $M_n=(1-2q)M_{n-1}$ so that the magnetization about every site decays
exponentially like $\vert M_n\vert =\vert M_0\vert \exp(-na/\xi )$.  The correlation length
$\xi = -a/\log \vert 2q-1\vert $ diverges only in the FM and AF limits and vanishes in the
NM state obtained by setting $q=1/2$.

In their earlier work on the Bethe lattice, Chattopadhyay {\em et al.} \cite{cha:00} assumed that the angles
$\theta_i$ were the same for every neighbor and inaccurately characterized this state by its
``incommensurate correlations'' rather than by its short-range order.  Since only the FM and AF states
possess well-defined wavevectors on the Bethe lattice, the SRO state cannot be interpreted using the formalism
originally developed by M\"uller-Hartmann \cite{mul:89} for the hypercubic lattice \cite{geo:96}.
Ordered spiral solutions are not allowed on the Bethe lattice due to its topology and lack of translational invariance.

Because the SRO phase has no well-defined wavevector, its transition temperature cannot
be obtained from the magnetic susceptibility \cite{fis:03,fis:05}.  Rather, $\TM (q)$
must be solved from coupled Green's function equations.  On a Bethe lattice, the bare and
full local Green's functions are related by
\begin{equation}
\uGr^{(\al )}_0(i\nu_n)^{-1}=z_n \uI -\frac{W^2}{16}\Bigl\{
q\, \uGr^{(\bal )}(i\nu_n) +(1-q)\, \uGr^{(\al )}(i\nu_n)\Bigr\},
\end{equation}
where $\uGr^{(\al )}$ and $\uGr^{(\bal )}$ are spin-reversed Green's functions and $z_n=i\nu_n +\mu $.
The full Green's function $\uGr^{(\al )} (i\nu_n)$ at site 0 is obtained using the effective action
of Eq.(\ref{act}).  With the parameterization
\begin{equation}
\label{g0}
\uGr^{(\al )}_0(i\nu_n )^{-1}=(z_n+R_n)\uI +Q_n^{(\al )}\underline{\sigma }_z,
\end{equation}
we evaluate the critical transition temperatures by linearizing in the magnetization
$M=\langle m_z\rangle $ on site $0$ (taken to be spin up).  To first order,
$R_n$ is independent of $M$ and is obtained by solving the cubic equation
\begin{equation}
\label{Rn}
\Bigl( (z_n+R_n)^2-\tJH^2\Bigr) R_n+\frac{W^2}{16}(z_n+R_n)=0,
\end{equation}
while $Q_n^{(\uparrow )}= -Q_n^{(\downarrow )}$ is proportional to $M$.
After integrating $\exp (-\seff (\hm ))$ over the Grassman variables, we find that the probability
for the local moment to point along $\hm $ is
$P(\hm ) \propto \exp (\beta J_{eff} M m_z )$ with
$J_{eff} =-2T\tJH \sum_n (Q_n^{(\uparrow )}/M)/\bigl( (z_n+R_n)^2-\tJH^2 \bigr)$.
Hence, $\TM (p,q)$ is solved from the implicit relation $\TM =J_{eff}(\TM )/3$ or
\begin{equation}
\label{TM}
1= -\frac{2\tJH^2}{3}(2q-1)\sum_n \frac{R_n}{(z_n+R_n)^2 (z_n+2(1-q)R_n)-\tJH^2(z_n+2(2-q)R_n/3)},
\end{equation}
which correctly reduces to the FM or AF results when $q=0$ or 1.  Notice that $\TM (p,q)=0$ in the NM state
with $q=1/2$.  Of course, $\TM (p,q)$ is the same for any arrangement of the spins
with the same average value of $\sin^2 (\theta_i/2) $.   In terms of the chemical potential $\mu $,
the electron filling $p$ is given by $p=1-(32T/W^2)\sum_n \re  R_n$.  So $p=1$ when $\mu =0$,
corresponding to half filling or one electron per site.

\begin{figure}
\includegraphics *[scale=0.6]{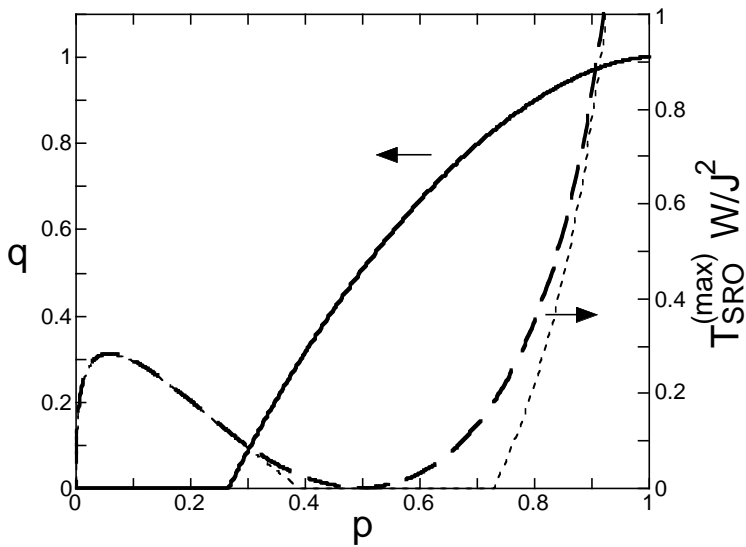}
\caption{
The correlation parameter $q$ (solid) and the associated maximum SRO transition temperature $\TR $ (dashed)
versus filling $p$ for a SRO state in the weak-coupling limit.  Also plotted in the thin short-dashed line are
the weak-coupling limits of $\TC $ for $q=0$ and $\TN $ for $q=1$.
}
\end{figure}

In the weak-coupling limit, $T \ll \tJH \ll W$, $\TM (p,q)$ is evaluated by converting the Matsubara
sum in Eq.(\ref{TM}) into an integral over $\nu $ and evaluating $R(z=i\nu +\mu )$ to zeroth
order in $\JH $.  Maximizing $\TM (p,q)$ with respect to $q$, we obtain $\TR (p)$
plotted in Fig.1.  Due to the symmetry about $p=1$, we restrict consideration to values of $p$
between 0 and 1.  Remarkably, a SRO phase with AF correlations ($1/2 < q < 1$) can always be found with
a higher transition temperature than the AF phase.  As $q\rightarrow 1$, the AF and SRO
transition temperatures meet at $p=1$ or half filling.  Even in the range of fillings
between about 0.26 and 0.38, where the Curie temperature is nonzero, a SRO phase with
FM correlations ($0 < q < 1/2$) has the higher critical temperature!  An instability is
also found in the range of fillings between 0.38 and 0.73, where neither the FM nor
AF phases are stable for small $\tJH /W$.  The point at which $q=1/2$ and $\TR (p) =0$
lies slightly below $p=0.5$.  The divergence of the weak-coupling results as
$p\rightarrow 1$ signals the breakdown of the condition $T\ll \tJH $ under which
they were derived and is associated with the appearance of a gap in the AF DOS, as discussed below.
The correlation parameter $q$ changes discontinuously at $\TR (p)$ from $1/2$ in the NM phase above to a
value less than or greater than $1/2$ in the SRO phase below.

We have also obtained the magnetic instabilities of the DE model for arbitrary $\tJH /W$.  For any $\tJH /W$ and $p$,
Fig.2(a) indicates the phase with the highest transition temperature.  Around $p=1$, an AF
instability occurs in a narrow range of fillings that vanishes as $\tJH /W \rightarrow 0$ or $\infty $.
The range of AF instabilities has a maximal extent (from $p=0.78$ to 1) when $\tJH /W \approx 0.33$.
For $\tJH /W < 0.33$, $\TR (p)=0$ when $q=1/2$ along the NM curve.
For $0.33 < \tJH /W < 0.5$, $\TR (p)>0$ for all fillings and $q$ jumps from 0
in the FM to a value in the range $1/2 < q < 1$ in the SRO phase.  For $\tJH /W > 0.5$,
either the FM or AF phase has a higher transition temperature than the SRO phase
and $q$ jumps from 0 to 1 at the FM/AF boundary.

\begin{figure}
\includegraphics *[scale=0.5]{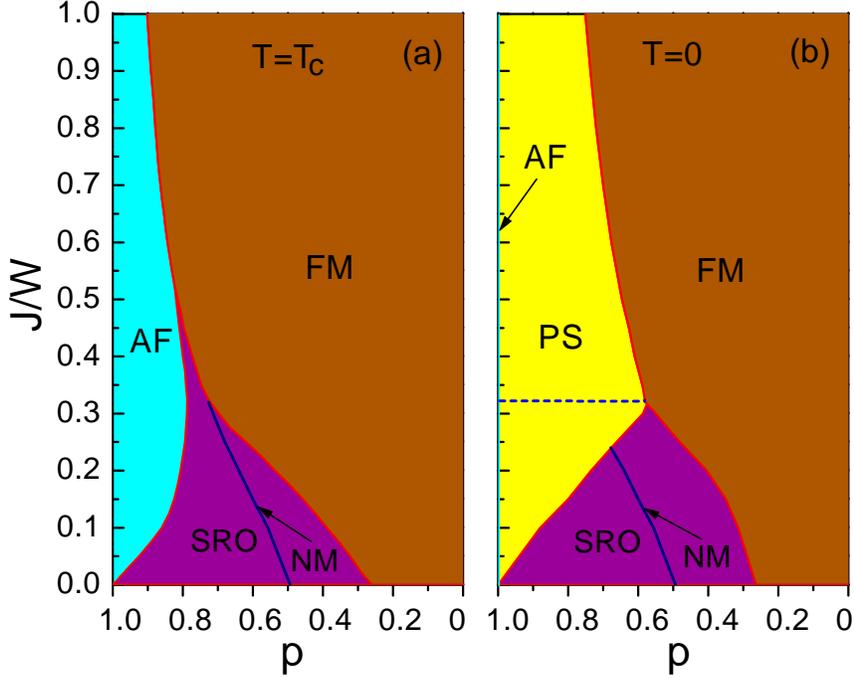}
\caption{
The (a) magnetic instabilities and (b) $T=0$ phase diagram of the DE model.
}
\end{figure}

The ground-state energies of the FM, AF, and SRO phases are obtained after
once again parameterizing $\uGr^{(\al )}_0(i\nu )^{-1}$ by Eq.(\ref{g0}).  Then $R(z)$
solves the quartic equation
\begin{equation}
\label{Rn4}
R(z+R)\Bigl\{ (z+2(1-q)R)^2-\tJH^2 \Bigr\} +\ds\frac{W^2}{16}(z+2(1-q)R)^2 = 0,
\end{equation}
which differs from Eq.(\ref{Rn}) (except when $q=1/2$) because we
are now working in the broken-symmetry phase.  The interacting DOS per spin is defined by
\begin{equation}
N(\omega ,q)=-\frac{1}{4\pi }\sum_{\al } \im \Ts \, \uGr^{(\al )}(z \rightarrow \omega +i\delta )
=\frac{16}{\pi W^2}\im \Bigl\{  R \Bigr\}_{z \rightarrow \omega +i\delta },
\end{equation}
which reduces to the bare DOS $N_0 (\epsilon )=(8/\pi W^2)\sqrt{W^2/4-\epsilon^2}$
when $\tJH \rightarrow 0$.  The AF DOS with $q=1$ can be solved analytically for all $\tJH $:
$N(\omega ,q=1) = (8\vert \omega \vert /\pi W^2)\re \sqrt{( W^2/4+\tJH^2 -\omega^2 )/(\omega^2 -\tJH^2 )}$,
which vanishes for $\vert \omega \vert  < \tJH $ and $\vert \omega \vert > \sqrt{W^2/4+\tJH^2}$.
Hence, $N(\omega ,q=1)$ contains a square-root singularity on either side of a gap with magnitude $2\tJH $.
As $\tJH /W \rightarrow \infty $, the width of each side-band narrows like $W^2/8\tJH $.
It is also possible to explicitly evaluate the DOS at $\omega =0$ for any $q$:
$N(\omega =0,q)=\bigl(8/\pi W^2(1-q)\bigr)\re \sqrt{ \tJH_c^2-\tJH^2 }$,
which vanishes when $\tJH > \tJH_c \equiv W(1-q)/2$.  In a NM, the critical value
required to split the band is $\tJH_c = W/4$, as found earlier \cite{che:03}.  In an AF,
$\tJH_c=0$ since a gap forms for any nonzero coupling constant.

Taking $\tJH /W=0.2$, we plot the interacting DOS versus $\omega $ for $q=0, 1/4, 1/2, 3/4,$
and 1 in Fig.3.  The FM DOS has kinks at $\omega =\pm (W/2-\tJH )$.  Between the
kinks, both up- and down-spin states appear;  on either side, the bands are
fully spin-polarized.  The kinks disappear in the SRO phase with $1 > q>0 $ due to the absence
of long-range magnetic order.  As $q$ increases from 0 to 1, the width of the DOS shrinks
and a gap appears when $\tJH > \tJH_c(q)$.  The breakdown of the
weak-coupling results in the limit $p\rightarrow 1$ is caused by the appearance
of a gap in the AF DOS for any nonzero $\tJH $.

\begin{figure}
\includegraphics *[scale=0.6]{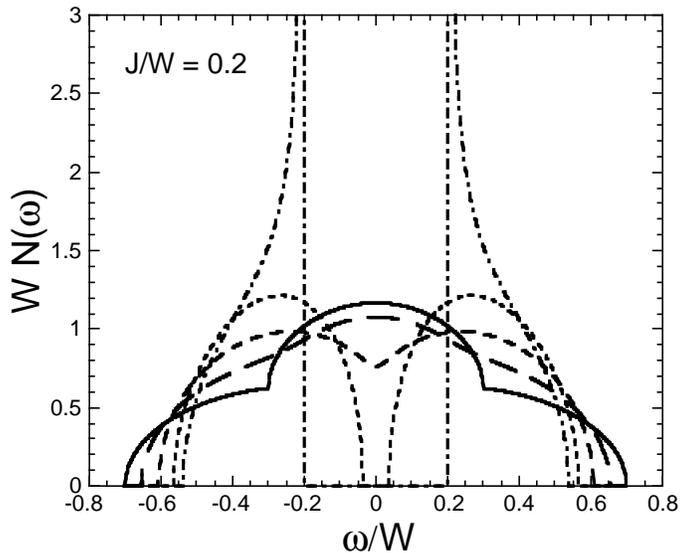}
\caption{
The interacting DOS $N(\omega ,q)$ (normalized by $1/W$) versus $\omega /W$
for $\tJH /W=0.2$ in the FM or $q=0$ (solid), SRO phase with $q=1/4$ (long dash),
NM or $q=1/2$ (dash), SRO phase with $q=3/4$ (small dash) and AF or
$q=1$ (dot dash) phases.
}
\end{figure}

Quite generally, the total energy $E(p,q)$ of any state may be written as an
integral over the interacting DOS:
$E(p,q)/N =2\int d\omega \, \omega f(\omega )N(\omega ,q)$,
where $f(\omega )=1/(\exp (\beta (\omega -\mu))+1)$ is the Fermi function and
$\mu =\mu (p)$.  At $T=0$, this relation becomes
\begin{equation}
\frac{1}{N}E(p,q)=-\frac{1}{\pi }\int d\nu \, \re \Biggl\{ 1 +\frac{16Rz}{W^2}\Biggr\},
\end{equation}
which converges because $R(z)\rightarrow -W^2/(16z)$ as $\vert z\vert \rightarrow \infty $.

In their numerical work, Chattopadhyay {\em et al.} \cite{cha:00} obtained a
very complex phase diagram at $T=0$, with SRO phases above $\tJH /W \approx 1/16$ and
PS between AF and FM phases for $0.1 < p < 1$ below this value.  In light of the numerical
difficulty of constructing the phase diagram for small $\tJH /W$, we have studied this limit
analytically.  As demonstrated below, PS disappears in the limit $\tJH /W \rightarrow 0$.

Carefully accounting for the dependence of the chemical potential
$\mu (p)$ on $\tJH /W$ for a fixed filling $p$, we find that the difference between the energies of the SRO and NM
phases \cite{He} is $\Delta E(p,q)/N =-3\TM (p,q)/2$, where both sides are evaluated analytically to second order in $\tJH /W$.
So for small $\tJH /W$, the ground state energy is minimized by the same correlation parameter
$q$ that maximizes the transition temperature!  This result is not surprising:  for small $\tJH /W$,
the transition temperature is also small so the correlation parameter that appears at $\TM (p,q)$
continues to minimize the energy at $T=0$.

It is now easy to understand why a SRO phase has a lower energy than the AF phase for small $\tJH /W$
when $p < 1$.  Due to the narrowing of the AF DOS, the AF energy may be higher than
that of a SRO phase with $\tJH < \tJH_c (q)$ and no energy gap.  When the chemical potential
of the AF lies outside the energy gap, then the square-root singularity of the AF
DOS will raise the energy compared to the broader, gapless DOS of a SRO state.

For any filling $p$, the energy $\Etmin (p)$ is obtained by minimizing $E(p,q)$ with
respect to $q$.  When $\tJH > 0$, PS occurs due to the formation of an energy gap
in the AF phase, so that $(1/N)d\Etmin /dp\vert_{p=1^-} =-\tJH $.
A necessary condition for PS between fillings $p_1 < 1$ and $p_2 =1$
is that the second derivative of the energy $\Etmin (p)$ must change sign at some filling $p^{\star }$
between $p_1$ and $p_2$.  To second order in $\tJH /W$, the
condition $d^2\Etmin (p)/dp^2=0$ may be written $1 /\sqrt{1-(2\mu /W)^2} \propto (\tJH /\mu )^2 $.
Since the right-hand side vanishes as $\JH \rightarrow 0$, the condition for PS cannot be satisfied
except as $\mu \rightarrow 0$ or $p^{\star }\rightarrow 1$.  Near $p=1$, $\mu /W \propto 1-p$ so
the PS width $p_2-p_1$ grows linearly with $\tJH/W$.

The magnetic instabilities and $T=0$ phase diagram of the DE model are plotted side-by-side in Fig.2.
For $\tJH /W < 0.33$, a SRO phase is stable over a range of fillings that diminishes with
increasing $\tJH /W$.  The NM solid curve in the SRO region with $q=1/2$ separates
a SRO phase with AF correlations ($1/2 < q <1$) on the left from one with FM correlations ($0 < q < 1/2$)
on the right.  For $\tJH /W < 0.33$, PS occurs between a SRO state and
an AF with $p=1$.  In agreement with the discussion above, the PS region grows linearly with
$\tJH /W$.  For $\tJH /W > 0.33$, the SRO phase is bypassed and PS is found
between a FM and an AF with $p=1$.  The horizontal dashed line in Fig.2(b) separates these two PS regions.
The PS width diminishes as $\tJH /W$ increases past 0.33.  In this high-coupling regime, our
results are in qualitative agreement with other authors \cite{fur:95,yun:98}.  Aside from the behavior at
very small $\tJH /W $, Fig.2(b) also agrees well with the phase diagram in Ref.\cite{cha:00}.

Together, Figs.2(a) and (b) provide a bird's eye view of the magnetic instabilities
and phase evolution in the DE model.  As the temperature is lowered, the phase space of the
pure FM shrinks and the AF phase with $p < 1$ disappears altogether.
The coupling $\tJH /W \approx 0.33$ below which a SRO phase is stable at
$T=0$ closely agrees with the value where the NM curve intersects the FM instability boundary
and where the AF instability covers the widest range of fillings in Fig.2(a).
Bear in mind that the pure FM, AF, and SRO phase instabilities of Fig.2(a) may be bypassed
when $\{ \tJH /W,p\}$ lies in the PS region of Fig.2(b).

The presence of a NM or SRO phase at $T=0$ would seem to contradict Nernst's theorem, which
requires that the entropy is quenched at zero temperature.  However, entropic
ground states occur quite frequently in infinite dimensions, appearing in earlier
work using a hypercubic lattice for the frustrated Hubbard \cite{geo:96} and
DE \cite{nag:00} models.

As pointed out over 40 years ago \cite{van:62}, the competing FM and AF interactions
in the DE model produce a RKKY-like interaction for small $J/W$ \cite{aus:01}, which
frustrates the ordered phases and stabilizes a SRO state.  For small $J/W$, the
component of the electronic spin along the local-moment direction is of order $J/W$ so the electronic
spins are not frozen at $T=0$.  When the dimension is lowered, the SRO state evolves into
the state with incommensurate correlations (IC) obtained by Monte Carlo simulations \cite{yun:98}.
Indeed, simply replacing ``SRO'' by ``IC,'' Fig.2(b) bears a striking resemblance to the one- and
two-dimensional phase diagrams of Ref.\cite{yun:98}.

Whether the SRO phase is a new kind of spin glass or spin liquid can only be resolved by future studies.
Two of the most important unresolved questions about spin glasses \cite{bin:86} are
whether there exists a true thermodynamic transition \cite{bit:97} and whether a model
without quenched disorder can support a spin glass \cite{sch:00}.   The answers for the SRO
phase of the DE model on a Bethe lattice are clear.  Although not marked by a divergent susceptibility,
the SRO transition is characterized by the development of short-range magnetic order and a
reduction in the entropy compared to the NM state.  In future work, we will pursue the analogy with
the spin-glass solution for the random Ising model on a Bethe lattice, where the Edwards-Anderson
order parameter can be explicitly constructed \cite{spgbl}.

To summarize, we have studied the behavior of a SRO solution to the DE model
in infinite dimensions.  Remarkably, the SRO transition temperature may be
higher than that of the ordered FM and AF phases and the SRO phase remains stable for small
couplings down to $T=0$.   The possibility of analytically constructing a disordered solution for the DE
model in infinite dimensions should be a boon to future investigations of the DE model in lower
dimensions.

It is a pleasure to acknowledge helpful conversations with Profs. Ian Affleck,
Elbio Dagotto, Jim Freericks, Mark Jarrell, Douglas Scalapino, and Seiji Yunoki.
This research was sponsored by the U.S. Department of Energy under contract
DE-AC05-00OR22725 with Oak Ridge National Laboratory, managed by UT-Battelle, LLC.
and by the National Science Foundation under Grant No. EPS-0132289 (ND EPSCOR).


\begin{references}
\bibitem{He} The same relation would be obtained for a Heisenberg model with classical spins
in mean-field theory (which becomes exact in infinite dimensions), so that
$H=-(1/2)\sum_{i,j}J_{ij} \vS_i \cdot \vS_j $ and $\Delta E /N = -J(\vk =0)S^2/2 = -3\TC /2$.
\end{references}
\end{document}